\documentclass[prb,letterpaper,twocolumn,showpacs,superscriptaddress]{revtex4-1}
\usepackage{hyperref}
\usepackage{amssymb}
\usepackage{amsmath}
\usepackage{float}
\usepackage{bbm}
\usepackage{graphicx}
\usepackage{epsfig}
\usepackage{epstopdf}
\usepackage[usenames]{color}
\def\x{{\bf x}}
\def\y{{\bf y}}
\newcommand{\Tr}{\text{Tr}}
\newcommand{\Slight}{{\ensuremath{\Sigma_{|0\rangle}}}}

\newcommand{\Sheavy}{{\ensuremath{\Sigma_{|\uparrow\downarrow\rangle}}}}
\newcommand{\Gup}{\ensuremath{G_{|\uparrow\rangle}}}
\newcommand{\Gdn}{\ensuremath{G_{|\downarrow\rangle}}}
\newcommand{\Gheavy}{\ensuremath{G_{|\uparrow\downarrow\rangle}}}
\newcommand{\Glight}{\ensuremath{G_{|0\rangle}}}

\begin{document}

\author{Emanuel Gull}
\affiliation{Department of Physics, Columbia University, New York, New York 10027, USA}
\author{David R. Reichman}
\affiliation{Department of Chemistry, Columbia University, New York, New York 10027, USA}
\author{Andrew J. Millis}
\affiliation{Department of Physics, Columbia University, New York, New York 10027, USA}

\title{Bold Line Diagrammatic Monte Carlo Method: General formulation and application to expansion around  the Non-Crossing Approximation}

\date{\today}

\hyphenation{}

\begin{abstract}
We  present a general framework for performing  ``bold-line'' diagrammatic Monte Carlo calculations using an analytical partial resummation as a starting point for a  stochastic summation of all diagrams. 
As a stringent test case we assess the accuracy of the method by solving the equations of single-site dynamical mean-field theory, using the non-crossing approximation as a starting point. We establish the validity of the starting approximations and show that the bold method provides a very accurate treatment of the Mott-insulating phase.
\end{abstract}

\maketitle

\section{Introduction \label{Introduction}}
Observables in quantum field theories may be expressed as infinite sums of Feynman diagrams.  During the last decade it has been realized that stochastic methods may be used to estimate the entire diagrammatic series.\cite{Prokofev96,Prokofev98} This requires a diagram-generating procedure which is ergodic (all diagrams contributing to the series must be generated) and which is such that each diagram is generated with a probability proportional to its weight in the series to be studied. These problems have been solved \cite{Prokofev98,Rubtsov05,Werner06,Gull08_ctaux,Werner09} and stochastic diagrammatic methods currently provide the best estimates for properties of the moderately correlated regime of the two-dimensional Hubbard model as well as successful impurity solvers for dynamical mean-field (DMFT)\cite{Georges96} and non-equilibrium \cite{Muhlbacher08,Werner09} problems.

Analytical studies over many years have established that while direct evaluation of low-order terms in a perturbation theory is rarely reliable, partial resummations of infinite series of diagrams often capture much of the relevant physics. Partial resummation replaces bare propagators (typically denoted by light lines in diagrams) with renormalized propagators (typically denoted by heavy or bold-face lines); thus, the diagrams involving further corrections to a partial resummation are often referred to as bold-line diagrams. Given these successes, it is natural to ask if stochastic techniques can be used to estimate the further corrections to an infinite partial resummation or, in other words, to develop a bold-line numerics.  References~\onlinecite{Prokofev08} and \onlinecite{Prokofev07} present one such method, a Monte Carlo computation based on the expansion of the self-energy of the polaron problem (single particle coupled to an oscillator bath) about the ladder resummation; however, a bold-line method for a truly many-body problem has heretofore not existed. 

In this paper we present a method for the  stochastic evaluation of a many-body fermionic bold-line perturbation theory. Our method is applicable to any diagrammatic series expansion.  We observe that in bare diagrammatic expansions,  different possible contractions for fermion operators typically sum up to determinants, substantially reducing the number of diagrams to be evaluated and ameliorating any minus sign problem. In bold expansions the determinant structure is lost and the important question is whether the loss is offset by the physics gained from the partial resummation. As a nontrivial example we apply the method to the one-impurity Anderson model, for which  analytical  resummation techniques are well established and are believed to capture much of the physics, and a substantial body of numerical work exists for comparison. The analytical resummation techniques are the non-crossing approximation\cite{Keiter71,Grewe81} (``NCA'')   and the one-crossing approximation\cite{Pruschke89} (``OCA''), where the names refer to specific topological features of the diagrammatics which we discuss in more detail below.   We formulate the bold-line expansion about these analytical resummations, delineate the regimes in which it is useful, and use the techniques to resolve a long-standing question concerning the form of the electronic spectral function near the edge of the Mott gap. As a by-product we determine the range of applicability of the NCA and OCA approximations for spin degeneracy $N=2$. 

\begin{figure}[tbh]
\begin{tabular}{cccc}
$a)$&\includegraphics[width=0.45\columnwidth]{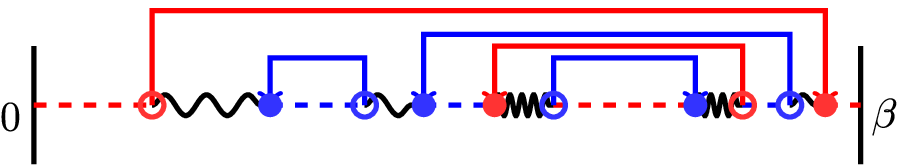}&
$c)$&\includegraphics[width=0.45\columnwidth]{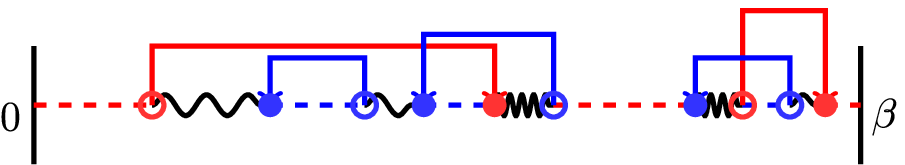}\\
$b)$&\includegraphics[width=0.45\columnwidth]{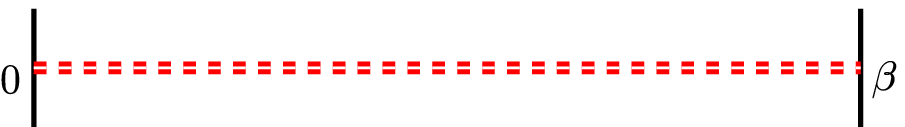}&
$d)$&\includegraphics[width=0.45\columnwidth]{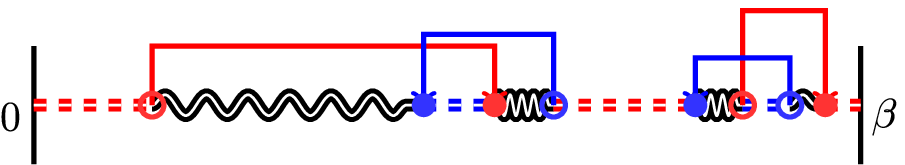}
\end{tabular}
\caption{(Color online) Typical diagrams arising in the hybridization expansion of the partition function of the Anderson impurity model: the four local states are described by wavy lines ($|0\rangle, |\uparrow\downarrow\rangle$), light, red (dark, blue) dashed lines ($|\uparrow\rangle, |\downarrow\rangle$). Light, red (dark, blue) solid lines denote hybridization functions, and empty (filled) circles local annihilation (creation) operators.
$a$: NCA diagram. $b$: ``bold'' propagator (double dashed line) which resums NCA diagrams including $a$. $c$: diagram with crossing lines. $d$: Diagram in expansion which resums diagrams including $c$. In BoldOCA, all of diagram $c$ is contained in the single bold line $b$.}
\label{diagrams}
\end{figure}

The rest of this paper is organized as follows: in Sec.~\ref{Methods} we outline the general features of our bold-line method. In Sec.~\ref{app} we present the specific formulas needed for the application of the method to the single-impurity Anderson model. In Sec.~\ref{Metrics} we  present metrics demonstrating the successes and limitations  of the method and in Sec.~\ref{Results} we show results for the electron Green's function, demonstrating in particular that the greatly improved accuracy of the method allows a definitive statement about the structure of the spectral function at the edge of the Mott-Hubbard gap. Section \ref{Conclusions} is a summary and discussion of future prospects. 

\section{Methods}\label{Methods}
Bold-line expansions of many-body problems require a modification of the diagrammatic Monte Carlo technique (these modifications were not necessary in the single-particle problem studied in Refs.~\onlinecite{Prokofev08} and \onlinecite{Prokofev07}). Conventional diagrammatic Monte Carlo is typically formulated in the configuration space consisting  of all diagrams contributing to the partition function. We refer to this space as partition function space and denote it by  $C_Z$. In a conventional expansion the space $C_Z$ suffices because Green's-function diagrams are generated by breaking a propagator line in partition function diagrams.  However, in a bold-line expansion the configuration space must be extended because there exist Green's-function diagrams which are not obtained by breaking lines  in a bold expansion of the partition function. We therefore employ a  ``Worm'' algorithm.\cite{Prokofev98,Burovski06} Our calculation is formulated in an extended configuration space $C_W$ consisting of the union of partition function ($C_Z$) and Green's function ($C_G$) space: $C_W = C_Z \cup C_G$, with a weight $w(\x)$ of a given configuration $\x$ given by $w_Z$ if $\x \in C_Z$ and $w_G$ if $\x \in C_G$. In this extended space we define a new partition function
\begin{align}
W = Z + \eta \int G,
\end{align}
where $\int$ denotes a sum over all the elements of the $G$ matrix and the parameter  $\eta$ is in principle arbitrary, but in practice should be chosen so that the $Z$ and $G$ parts of $W$ give comparable contributions; the calculations presented here use $\eta=0.15.$

A measurement of a component $ab$ of $G$, $\langle G_{ab}\rangle_W$, in  the extended space $C_W$ is proportional to the physical Green's function but is wrongly normalized. The correct normalization is obtained after division by the partition function, also measured in the space $C_W$, i.e. by $\langle \delta^Z\rangle_W = \sum_\x \mathfrak{z}_\x$ with $\mathfrak{z}_\x=1$ if $\x \in C_Z$ and $\mathfrak{z}_\x=0$ otherwise.
Thus,
\begin{align}
\langle G_{ab}\rangle = \frac{\langle G_{ab}\rangle_W}{\langle \delta^Z\rangle_W}.
\end{align}

The sum over all terms in $\mathcal{C}_W$ is performed using a diagrammatic Monte Carlo method: diagrams in $\mathcal{C}_W$ are generated, accepted, or rejected stochastically by inserting and removing local operators and hybridization lines  according to their contribution to $Z_W$ and integrated stochastically, in analogy to Ref.~\cite{Werner06}. This summation is exact if all bold diagrams are included. 

While diagram-generating procedures are model dependent, they generically consist of the insertion or removal of operators and reconnection of parts of diagrams. The proposed insertion (removal) of operator tuples (typically pairs) raises (lowers) the diagram order, changing a configuration $\x$ to $\y$ and is accepted with probability
\begin{align}
W^\text{acc}_{\x\y} &= \min\big(1, R_{\x\y}\big),\\
R_{\x\y} &= \frac{w(\y)W^\text{prop}_{\y\x}}{w(\x)W^\text{prop}_{\x\y}} = R_{\y\x}^{-1},\label{MetropolisR}
\end{align}
where $W^\text{prop}_{\x\y}$ denotes the proposal probability of an update. Transitions between configurations in $\mathcal{C}_G$ and $\mathcal{C}_Z$ are performed by inserting or removing Green's function operators or propagator lines
into (from) a partition function configuration. Typical diagram generating procedures can produce terms which are already included in the bold resummation; one must test each generated diagram to make sure that it is not already included, but the computational cost of the test is typically negligible.  

\section{Application: Single-Impurity Anderson Model}\label{app}

As a stringent test of the bold-line resummation methods we consider one of the best-studied nontrivial models in condensed-matter physics, namely, the single-impurity Anderson model. This model represents the physics of a magnetic impurity in a metal and is also important as an auxiliary problem in the ``dynamical mean-field'' approximation to the properties of models of correlated electron materials.\cite{Georges96} For this model, widely used partial resummation methods are available.  In addition, numerically exact results  obtained by the hybridization expansion (CT-HYB) quantum Monte Carlo method \cite{Werner06}  are available for comparison.

The Anderson model describes a correlated site coupled to a bath of free electrons; the Hamiltonian is
\begin{equation}
H = H_\text{bath} + H_\text{mix} + H_\text{loc}
\label{HAnderson}
\end{equation}
with
\begin{align*}
H_\text{bath} &=\sum_{p\sigma}\epsilon_p a_{p\sigma}^\dagger a_{p\sigma};\ \  H_\text{mix} = \sum_{p\sigma}(V_{p} a_{p\sigma}c_{\sigma}^\dagger+ h.c.);\\H_\text{loc} &= \sum_{\sigma}-\mu n_{\sigma}+U n_\uparrow n_\downarrow,
\end{align*}
where the $a_{p\sigma}$ label a continuum of ``bath'' operators with dispersion $\epsilon_p$, and $c_\sigma$ are local operators at energy $\mu$ with interactions $U$, hybridizing with the bath with strength $V_{p}.$  In the original formulation of the Anderson model, the parameters $\varepsilon_d$ and $U$ are properties of the magnetic impurity and $\varepsilon_p$ and $V_p$ are properties of the host metal. In the dynamical mean-field context the impurity model is an auxiliary problem used to provide information about a lattice model of interacting electrons. In this case $U$ is the on-site interaction of the lattice model and the $\varepsilon_d,\varepsilon_p$ and $V_p$ are fixed by a self-consistency condition, as discussed  in Ref. \onlinecite{Georges96}.    These details are not important for the formulation and application of the bold-line expansion. However, our specific results are obtained using parameters arising from the dynamical mean-field solution of the one-orbital Hubbard model at various particle densities and interaction strengths. 

We study the model as an expansion in $V$ about the atomic limit in which the local states are decoupled from the bath.  Alternatively, a resummation of interaction diagrams around the free limit could be designed; we do not consider this here. The expansions may be represented pictorially by time-ordered diagrams such as those shown in Fig.~[\ref{diagrams}].  The propagation in atomic states is represented by different lines (wavy and dashed): in the eigenbasis $|0\rangle, |\uparrow\rangle, |\downarrow\rangle,|\uparrow\downarrow\rangle$ of $H_\text{loc}$ the energies are $E_\text{loc}=0, -\mu, -\mu, U-2\mu,$ respectively, and the corresponding (bare) propagators are $e^{-\tau E_\text{loc}}$. The hybridization vertices are indicated by solid and open circles while  the propagation of electrons in the bath is denoted  by solid lines which represent the hybridization function
\begin{equation}
\Delta(\tau) = \sum_p \frac{V^{*}_p V_p}{e^{-\epsilon_p\beta}+1} \times \left\{ \begin{array}{ll}-e^{\epsilon_p(\tau-\beta)},& \tau > 0 \\ e^{\epsilon_p\tau},& \tau < 0.\end{array}\right.
\end{equation}  
Accurate and efficient numerical methods exist for evaluating the direct hybridization expansion.\cite{Werner06} 

The resummation methods we consider are the NCA (Refs.~\onlinecite{Keiter71} and \onlinecite{Grewe81}) and the  OCA.\cite{Pruschke89} The NCA resums all segments containing no crossing fermion lines [e.g., Fig.~\ref{diagrams} (a)] into a renormalization of the propagator of the atomic state [e.g., Fig.~\ref{diagrams} ($b$)], which becomes $G_{|j\rangle} = [(G_{|j\rangle}^0)^{-1}-\Sigma_{|j\rangle}]^{-1}$ with $\Sigma_{|j\rangle}$ given by the self-consistent equations
\begin{align}
\Slight(\tau) &= \Gup(\tau)\Delta_\uparrow(\tau) + \Gdn(\tau)\Delta_\downarrow(\tau),\\
\Sigma_{|\sigma\rangle}(\tau) &= \Glight(\tau)\Delta_\sigma(-\tau) + \Gheavy(\tau)\Delta_{-\sigma}(\tau),\\
\Sheavy(\tau) &= \Gup(\tau)\Delta_\downarrow(-\tau) + \Gdn(\tau)\Delta_\uparrow(\tau).
\end{align}
First-order OCA self-energies are given by equations involving additional crossing hybridization lines, and the full OCA equations are obtained by resumming vertex equations. Note that the projection techniques used in analytical  NCA calculations \cite{Bickers87,Hewson93} are not needed here. Multi-orbital \cite{Werner06Kondo} and cluster models can be studied using the same operations with $\Delta$, $G$, and $\Sigma$ in matrix form.

The hybridization expansion of the partition function is
\begin{align}\label{PFExp}
Z &= \sum_k \iiint d\tau_1 \cdots d\tau_k' \sum_{j_1, \cdots j_k \atop j_1', \cdots j_k'}\det \Delta\\ \nonumber
&\times\Tr_c\left[ e^{-\beta H_\text{loc}} T_\tau c_{j_k}(\tau_k) c_{j_k'}^\dagger(\tau_k')\cdots c_{j_1}(\tau_1)c_{j_1'}^\dagger(\tau_1')\right].
\label{FiniteTPFExp}
\end{align}
The hybridization expansion algorithm \cite{Werner06,Werner06Kondo} (CT-HYB) enables a direct numerical sampling of this series. To sample a bold-line expansion around the NCA we restrict to diagrams that contain no non-crossing parts (i.e., we use the CT-HYB method to generate diagrams but  do not sample those [such as Fig.~\ref{diagrams}($c$)] that have a part contained in NCA, and we replace the atomic propagators $G^0_{|j\rangle}$ by NCA propagators $G^{NCA}_{|j\rangle}$). In order to sample around the first-order OCA, we limit ourselves to diagrams without one-crossing subsegments [thereby reducing Fig.~\ref{diagrams}($c$) to Fig.~\ref{diagrams}($b$)].  Testing if a diagram needs to be sampled is a (cheap) $O(k)$ operation: in BoldNCA diagrams with hybridization lines $a_p^\dagger a_p$ that span no other local operators need not be sampled. In BoldOCA, additionally, diagrams of the type $a_p^\dagger a_q a_p a_q^\dagger$ are excluded.

The CT-HYB diagram-generating procedure is ergodic (generates all diagrams); while we do not have a formal proof that it remains ergodic when restricted to bold diagrams we have been unable to find a counterexample and  have extensive numerical evidence that  all bold diagrams are generated.

An independent discussion  of a bold expansion about the NCA limit was given in Ref.~\onlinecite{HauleBold}. The extended configuration space $C_W$ was not employed. A bold expansion as defined here was not implemented but an evaluation of all diagrams up to 5th order, with time integrals evaluated by a Monte Carlo method, was reported.

\section{Results: Metrics}\label{Metrics}

\begin{figure}[tbh]
\includegraphics[width=0.8\columnwidth]{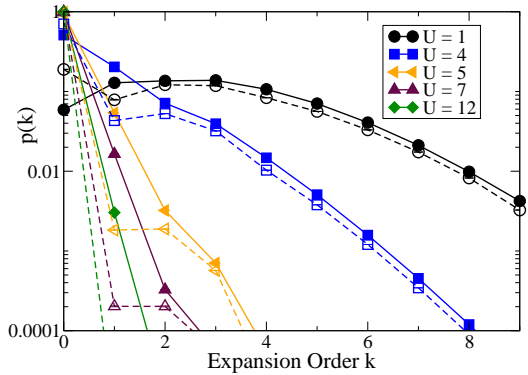}
\caption{
Probability $p$ for a diagram of $Z$ to contain $k$ spin-up hybridization lines, computed for Anderson impurity model with semicircular density of states and hybridization function fixed by dynamical mean-field self-consistency condition with parameters $\eta=0.15$ $n=1, \beta t=10$, at $U$-values indicated. Full symbols: BoldNCA. Empty symbols: BoldOCA..
}
\label{order_nca}
\end{figure}

In this section we present  some basic results  which illustrate the strengths and weaknesses of the bold expansion. Our results were obtained using the Anderson model corresponding to the dynamical mean-field approximation to the single-orbital Hubbard model on a Bethe lattice.\cite{Imada98,Georges96} This model is specified by  an interaction strength $U$, a hopping parameter $t$, and a carrier concentration $n$.  At carrier concentration $n=1$ the dynamical mean-field approximation to this  model has a metal-insulator transition at a critical interaction strength $U\approx 5t$ (at the temperatures we study) \cite{Georges96} and we shall see that the bold expansion behaves very differently in the insulating and metallic phases.   

An important metric is the weight $p(k)$ of terms at order $k$ in the expansion for $Z$. This is shown in Fig.~\ref{order_nca} for different interaction strengths $U$ at  inverse temperature $\beta =10/t$. The curves are characterized by a non-negligible weight at zero expansion order and a tail which decays approximately exponentially at high perturbation order.  If the NCA (or first-order OCA) exactly solved the model $p(k=0)$ would equal $1$ and the tail would be absent. 

A clear difference between the curves is seen. For $U\gtrsim 5$ $p(k=0)$ is very close to unity and the weight in the tail is very small, while for $U\lesssim 5t$ $p(k=0)$ becomes smaller and the contribution of the non-NCA (non-OCA) diagrams becomes important.  These differences arise from a difference in physics. At this temperature the model is in a gapped insulating phase for $U \gtrsim 4.5t$  and is in a gapless metallic phase for $U\lesssim 4.5t$. Clearly the NCA and first-order OCA are very good approximations to the insulating phase and poor approximations to the metallic phase.  A bold expansion around the free ($U=0$) limit (not considered here) presumably behaves differently: while the weakly correlated metal would be captured accurately the local physics of the Mott insulator would be difficult to reach.

\begin{figure}[tbh]
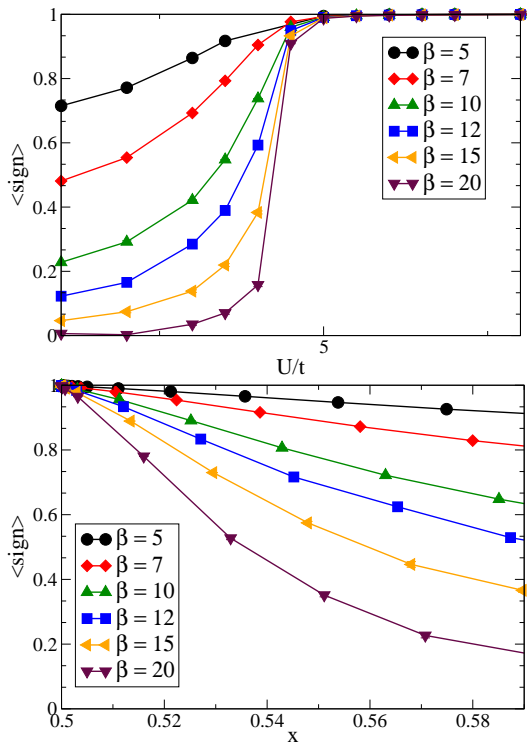

\includegraphics[width=0.8\columnwidth]{sign_OCA.eps}\\
\includegraphics[width=0.8\columnwidth]{sign_OCA_doping.eps}
\caption{
Upper panel: expectation (over $C_W$) of sign as a function of interaction $U$, at carrier concentration $n=1$ per site and temperatures indicated.  Lower panel: expectation value of sign as a function of carrier concentration $1-x$ at $U=6t$ and temperatures indicated.
}
\label{sign_nca}
\end{figure}

We have also studied these histograms of $p(k)$ as a function of temperature (not shown). The temperature dependence is negligible in the insulating phase. In the metallic phase, as the temperature is decreased, the value $p(k=0)$ decreases and the weight in the tail increases; as $T\rightarrow 0$ $p(k=0)\rightarrow 0$ reflecting the failure of NCA and OCA to adequately describe the physics in the $T\rightarrow 0$ limit of the metallic phase. Note that the weight of diagrams as a function of expansion order as expressed by the long tail in Fig.~\ref{sign_nca} only decays very slowly and that basing the algorithm on a more sophisticated resummation (here, first-order OCA instead of NCA) does not change the decay of the tail. 

Another important metric for an expansion of an interacting fermion problem is the average sign. Figure~\ref{sign_nca} shows that the bold expansion suffers from a sign problem.  The upper panel demonstrates that in the insulating regime ($U\gtrsim 4.5$, at this $T$)  the expansion around BoldNCA shows convergence at very low order and configurations with negative sign (which occur at higher expansion order)  give a negligible contribution. Thus, in this case the loss of the determinant structure is compensated by the much better starting solution. However, in the metallic case the starting solution is less good and a severe sign problem arises. The steep drop around $U/t=4$ marks the departure from the insulator and the failure of NCA and OCA. The lower panel shows a similar behavior of the sign as a function of doping at a strong interaction.

\begin{figure}[tbh]
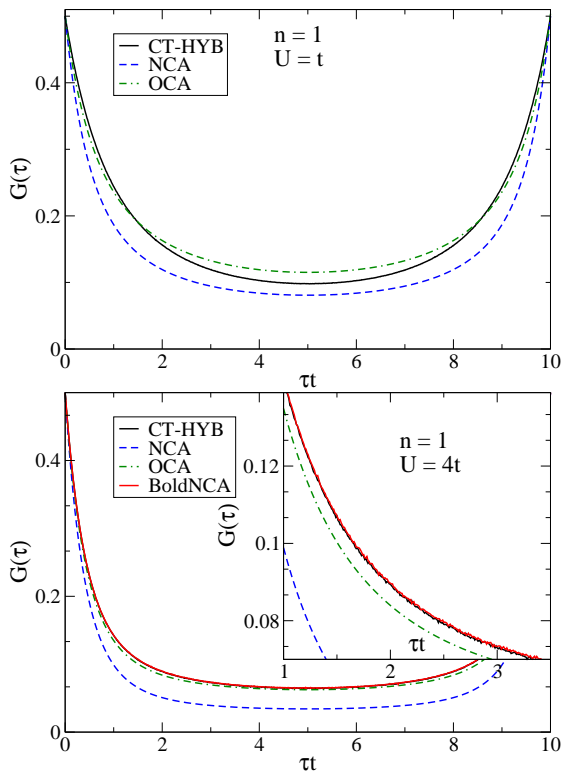

\includegraphics[width=0.85\columnwidth]{GreenNCAOCAMetal.eps}\\
\includegraphics[width=0.85\columnwidth]{GreenNCAOCACorrMetal.eps}
\caption{Green's function of Anderson impurity model with semicircular density of states and hybridization function fixed by a DMFT self-consistency condition, calculated using NCA, OCA, CT-HYB and BoldNCA (bold OCA would be indistinguishable), $\beta t=10$, starting from converged and accurate \cite{Gull07} CT-HYB results, at interactions and dopings indicated.
Upper panel: Fermi liquid case $U/t=1$. Lower panel: Correlated Metal $U/t=4$.
\label{green1}
}
\end{figure}

\begin{figure}
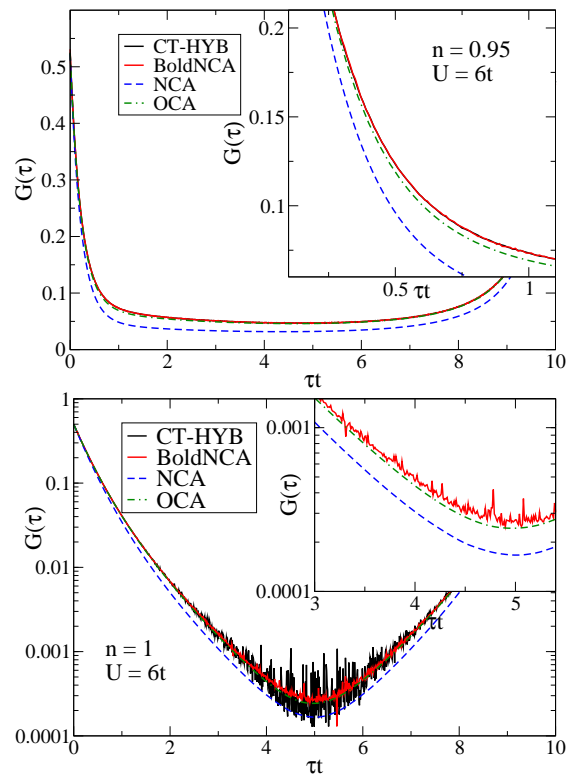

\includegraphics[width=0.85\columnwidth]{GreenNCAOCA_awayHalfFilling.eps}\\
\includegraphics[width=0.85\columnwidth]{GreenNCAOCA.eps}
\caption{Green's function of Anderson impurity model with semicircular density of states and hybridization function fixed by a DMFT self-consistency condition, calculated using NCA, OCA, CT-HYB and BoldNCA (bold OCA would be indistinguishable), $\beta t=10$, starting from converged and accurate \cite{Gull07} CT-HYB results, at interactions and dopings indicated.
Upper panel: Doped Mott insulator at filling $n=0.95$ and interaction $U/t=6$. Lower panel: Mott insulator, $U/t=6$, $n=1$.
}
\label{green2}
\end{figure}

\section{Results: electron Green's function}\label{Results}

Figures \ref{green1} and \ref{green2} compare the imaginary time Green's function obtained by BoldNCA (BoldOCA results are indistinguishable on the scale used in Figs.~\ref{green1} and \ref{green2}) simulations to the CT-HYB results and analytical NCA and OCA results. Comparable computational resources are invested in the bold and CT-HYB calculations. The large differences between the NCA and OCA and the Monte Carlo curves in the upper panel of Fig.~\ref{green1} show that  for the weakly correlated metallic phase the initial starting point is poor and the bold expansion is not useful in practice. The lower panel shows that in the moderately correlated ``bad metal'' case the starting point is closer to the exact answer and the bold and CT-HYB results are very close at temperature $T=t/10$. The rapid decrease of sign with $T$ (cf. Fig.~\ref{sign_nca}) shows that for $U/t=4,$ $\beta t\approx15$ the bold method will become significantly less efficient. The upper panel of Fig.~\ref{green2} shows that the same situation is obtained in the strongly correlated, lightly doped case. Finally, the lower panel of Fig.~\ref{green2} demonstrates the clear superiority of the bold methods in the insulating case, where the noise in the center of the imaginary-time interval is very substantially reduced. 

\begin{figure}[tbh]
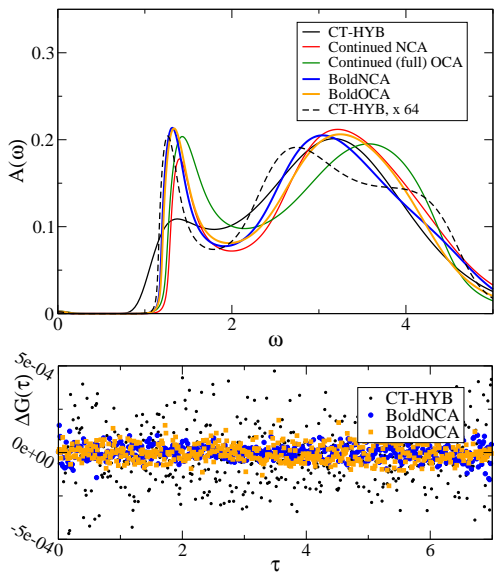

\includegraphics[width=0.75\columnwidth]{spectra.eps}\\
\includegraphics[width=0.75\columnwidth]{recont.eps}
\caption{Upper panel: Analytic continuation of Green's functions obtained from different solvers from DMFT of half-filled Hubbard model at $U=6t$ and $\beta t=7$. Lower panel: difference $\Delta G(\tau)$ between $G(\tau)$ obtained from CT-HYB, BoldNCA, and BoldOCA measurements and $G(\tau)$ reconstructed from the analytically continued $A(\omega)$.
}\label{analcont}
\end{figure}
In standard CT-HYB simulations, spectra of insulating systems are hard to obtain because of large (relative) errors in the mid-range of the imaginary-time interval and so the questions of the precise value of the insulating gap and the form of the above-gap structure have been discussed extensively in the literature; for recent work and references, see Refs.~\onlinecite{Wang09} and \onlinecite{Krivenko06}.   Figure \ref{analcont} shows spectral functions obtained by maximum entropy analytical continuation of the bold-line, CT-HYB, and analytical NCA and OCA approximations. The relative errors in the basic CT-HYB data lead to a substantial smearing of the gap edge features. Also shown is the continuation of high-precision CT-HYB data obtained by expending 64 times more computer resources.  The bold expansions and the high-precision CT-HYB data essentially agree on the gap value and the form of the spectral function near the gap edge. (The differences at higher frequency reflect the intrinsic sensitivity of analytical continuation to very small differences in data.) 

Our results firmly establish that the spectral function in the paramagnetic insulating phase is characterized by a sharp peak at the gap edge, and we obtain more precise values for the insulating gap. The accuracy is confirmed by the lower panel of Fig.~\ref{analcont} which presents the difference $\Delta G$ between the measured $G(\tau)$ and $G(\tau)$ back-continued from $A(\omega)$.  The analytical approximations show clear deviations, in particular a larger gap. The difference in gap value can be seen directly as a difference in the imaginary-time data in the lower panel of Fig.~\ref{green2}: $G(\tau)$ calculated using the two analytical methods falls below the numerically exact results in the imaginary-time range $2<\tau t<5$.  

\section{Conclusions \label{Conclusions}}
In conclusion, we have presented a ``bold'' diagram method and applied it to an expansion around the noncrossing and one crossing approximations to  the Anderson impurity model.  We have  also applied the algorithm to the Kondo limit of the Anderson model without a self-consistency condition, with results (not shown) very similar to those described here. The  NCA and OCA approximations  are believed to become exact in the limit of large $N$,\cite{Bickers87} but their accuracy for physically relevant $N$ has been established here. 
BoldNCA is general, numerically exact, and easily extensible to multiple orbitals or cluster calculations, where a severe sign problem appears also in the hybridization expansion and the effect of reducing the expansion order will be most pronounced. We expect the main region of applicability to be in this area, as well as to nonequilibrium problems, where the oscillating phase severely limits the applicability of non-bold methods.\cite{Werner09}  Extensions to non-equilibrium and multi-orbital calculations and to the resummation of vertices are currently under way. Diagrammatic random phase approximation (`RPA') resummation is also essential for the study of screening and polarization; exploration of the methods discussed here may be fruitful in this context.

\acknowledgments{We thank X.~Wang for providing continuations shown in Fig.~\ref{analcont}. E.~G. and A.~J.~M. are supported by NSF under Grant No.~DMR-0705847, and acknowledge partial support from NSF Grant No. PHY05-51164. Our programs are based on the ALPS  library.\cite{ALPS} Calculations were performed on the Brutus cluster at ETH Zurich.}
\bibliography{refs_shortened}
\end{document}